\documentclass[prc,showpacs,twocolumn,nofootinbib]{revtex4}
\usepackage{graphicx}
\usepackage{dcolumn}
\usepackage{bm}

\begin{document}

\title{Microscopic description of quadrupole-octupole coupling in
actinides  
with the Gogny-D1M energy density functional}

\author{R. Rodr\'{\i}guez-Guzm\'an}

\email{raynerrobertorodriguez@gmail.com}

\affiliation{Department of Physics, Kuwait University, Kuwait }

\author{Y. M. Humadi}

\affiliation{Department of Physics, Kuwait University, Kuwait }

\author{L.M. Robledo}

\affiliation{%
Center for Computational Simulation, Universidad Polit\'ecnica de 
Madrid, Campus Montegancedo, 28660 Boadilla del Monte, Madrid, Spain
}%

\affiliation{Departamento  de F\'{\i}sica Te\'orica and CIAFF, 
Universidad Aut\'onoma de Madrid, 28049-Madrid, Spain}

\email{luis.robledo@uam.es}

\date{\today}

\begin{abstract}
The interplay between quadrupole and octupole degrees of freedom is 
discussed in a series of U, Pu, Cm and Cf isotopes both at the 
mean-field level and beyond. In addition to the static 
Hartree-Fock-Bogoliubov approach, dynamical beyond-mean-field 
correlations are taken into account via both parity restoration and 
symmetry-conserving Generator Coordinate Method calculations based on 
the parametrization D1M of the Gogny energy density functional. 
Physical properties such as  correlation energies, negative-parity 
excitation energies as well as reduced transition probabilities $B(E1)$ 
and $B(E3)$ are discussed in detail and compared with the available 
experimental data.  It is shown that, for the studied nuclei, the 
quadrupole-octupole coupling is  weak and to a large extent the 
properties of negative parity states can be reasonably well described 
in terms of the octupole degree of freedom alone.
\end{abstract}


\maketitle{}

%
%
%

\section{Introduction.}
Fingerprints of octupole collectivity in even-even nuclei are usually 
associated with the presence of $1^{-}$ states in  the low-lying 
spectra. As the ground state of those nuclei is usually quadrupole 
deformed, there is a $3^{-}$ state, member of the corresponding 
negative-parity rotational bands, which decay through fast $E3$ 
transitions to the $0^+$ ground state. On the other hand, the $1^{-}$ 
state decays via $E1$ transitions. The exploration of these as well as 
other unusual features associated with octupole correlations 
\cite{Ahmad_93,butler_2016,butler_2015} already started in the 1980s 
\cite{butler_96} and has become an active field of research since then 
- see Refs 
\cite{Gaffney_2013,Tandel_2013,Li_2014,Ahmad_2015,Bucher_2016} for some 
recent examples. In \cite{Gaffney_2013} the measured $E3$ strength in  
$^{220}$Rn and $^{224}$Ra unambiguously established the octupole 
deformed character of the later nucleus. This represents the first 
unambiguous experimental evidence of permanent octupole deformed 
even-even nucleus. In multi-step Coulomb excitation experiments 
performed at the ATLAS-CARIBU facility with $\gamma$-ray and 
charged-particle detectors \cite{Bucher_2016} also large $E3$ 
transition strength in $^{144}$Ba was found pointing to a permanent 
octupole deformed ground-state. Evidence for permanent octupole 
deformation in $^{146}$Ba has subsequently been obtained 
\cite{Bucher_146Ba_2017}. Recent experiments \cite{Butler2020} have 
also established the octupole deformed character of $^{222}$Ra, or 
measured the $E1$ strength in $^{228}$Th \cite{Chishti20}.

\begin{figure*}
\includegraphics[width=0.9\textwidth]{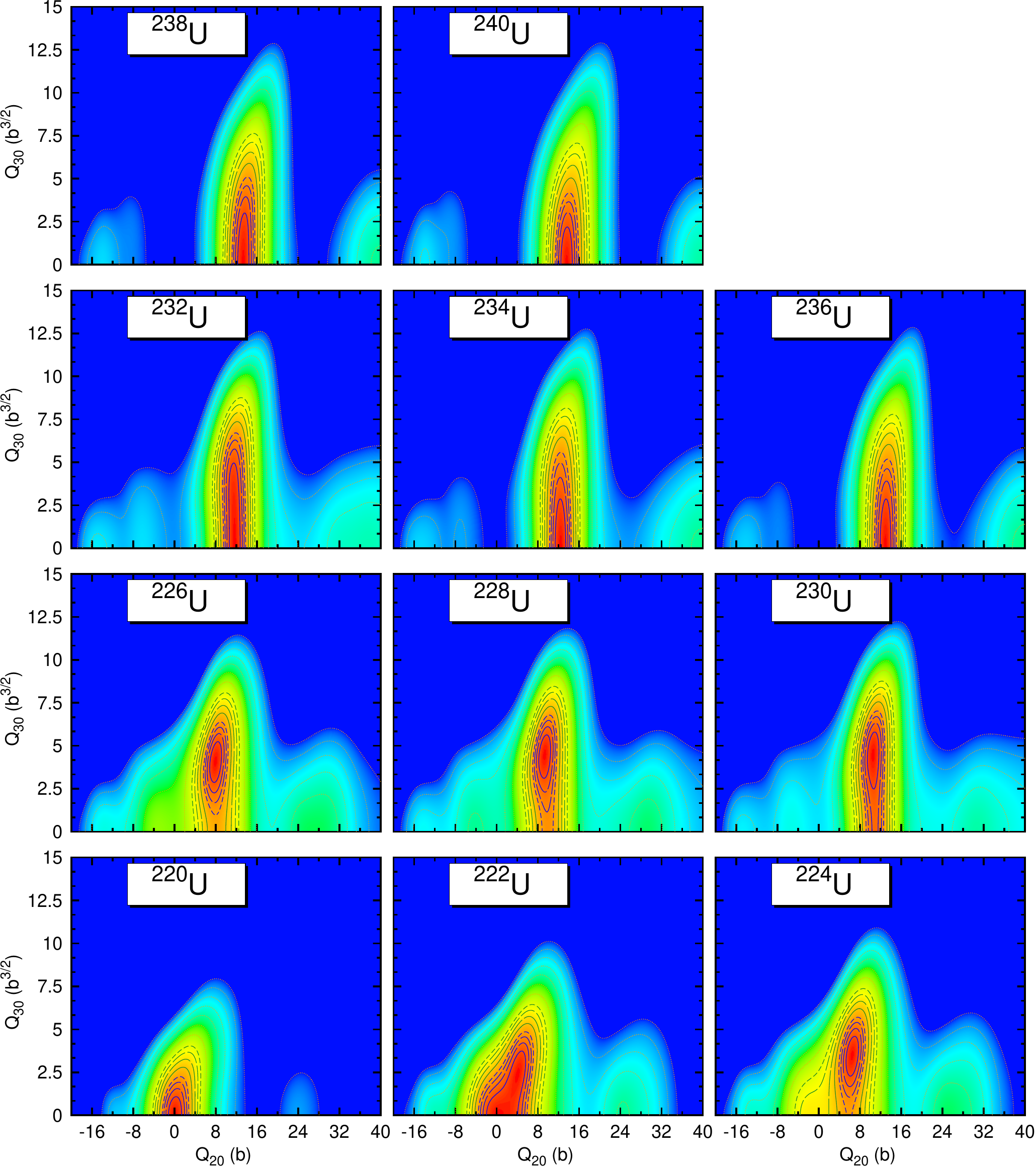}
\caption{(Color online) MFPESs computed with the Gogny-D1M EDF for the 
isotopes $^{220-240}$U. Taking the lowest mean-field energy as a 
reference, solid and dashed contour  lines extend  from 0.25 MeV  up to 
1 MeV  in steps of 0.25 MeV. Solid and dashed contours are then drawn 
in steps of 0.5 MeV up to 3 MeV and from there up dotted lines are 
drawn in steps of 1 MeV. The intrinsic HFB energies are symmetric under 
the exchange $Q_{30} \rightarrow -Q_{30}$. For $A=230$, the conversion
factor from barn to $\beta_2$ values is 0.0212 and the one from $b^{3/2}$ to
$\beta_3$ values is 0.0342. For additional details, see the 
main text.
}
\label{Q2Q3_MF_U} 
\end{figure*}

From a theoretical point of view, various techniques and models have 
been employed to study the dynamics of octupole collectivity 
\cite{butler_96,moller_81,leander_82,naza_84,naza_92,babilon_05,minkov_06}. 
Some of the approaches use potential energy surfaces (PESs) obtained within  
relativistic and nonrelativistic mean-field approximations to obtain 
the parameters of the  Interacting Boson Model 
\cite{Bing_2014,Nomura_Ba_RE_2018,Nomura_Th_RE_2018,Nomura_Rayner_IBM_2015}. 
Some others rely on microscopic frameworks, both at the mean-field 
level and beyond, based on the nonrelativistic Skyrme and Gogny as well 
as relativistic energy density functionals (EDFs) 
\cite{mar83,bon86,bon91,hee94,erler-85,Ebataba-2017,rob87,rob88,egi90,egi91,gar98,rob10,egi92,Fission-D1Mstarstar,Long_2004,Tomas_GCM_parity_2016,Xia_PRC_2017,Xu-2017,Agbemava-Q3-2016,Agbemava-Q3-2017,Recent-Survey-Q3}.

Octupole deformation properties of several even-even actinides  were 
discussed in Ref.~\cite{JPG_2012_RoRay} with the help of 
octupole-constrained  Hartree-Fock-Bogoliubov (HFB) calculations based 
on the parametrizations D1S \cite{gogny-d1s}, D1N \cite{gogny-d1n} and 
D1M \cite{gogny-d1m} of the Gogny \cite{gogny}  and the BCP 
\cite{BCP-1,BCP-2,BCP-3,BCP-4} EDFs. A one-dimensional (1D) collective 
Hamiltonian was also built to have access to properties such as the 
excitation energies of $1^{-}$ states as well as  $B(E1)$ and $B(E3)$ 
transition probabilities. A thorough account over a large set of 
even-even nuclei of observables associated to octupole correlations   
was presented in Refs.~\cite{Robledo-Bertsch-Q3-1,Robledo15} using the 
octupole-constrained Gogny-HFB approach, parity projection and octupole 
configuration mixing. From the results of these studies it is clear that 
not only static octupole deformation plays a role but also dynamical 
octupole correlations have a sizable impact on observables.

The interplay between quadrupole transitional properties and octupole 
deformation manifestations in a selected set of Sm and Gd nuclei was 
discussed in Ref.~\cite{Rayner_Q2Q3_GCM_2012} using the  D1S and D1M 
Gogny-EDFs. Both  quadrupole and octupole constrains were considered 
simultaneously. The mean-field potential energy surfaces (MFPESs) 
obtained for $^{146-154}$Sm and $^{148-156}$Gd exhibited a very soft 
behavior along the octupole direction indicating, that dynamical 
beyond-mean-field  effects should be taken into account. Those 
beyond-mean-field  effects were considered via both parity projection 
of the intrinsic states and symmetry-conserving quadrupole-octupole 
configuration mixing calculations, in the spirit of the two-dimensional 
(2D) Generator Coordinate Method (GCM) \cite{rs}. In addition to the 
systematic of the $1^{-}$ excitation energies, correlation energies, 
$B(E1)$ and $B(E3)$ transition probabilities, the results of 
Ref.~\cite{Rayner_Q2Q3_GCM_2012} suggested a shape/phase transition 
from weakly to well quadrupole deformed ground states as well as a 
transition to an octupole vibrational regime in the studied nuclei. The 
quadrupole-octupole coupling has also been studied for Rn, Ra and Th 
nuclei within the 2D-GCM framework \cite{Robledo_2D-GCM_with_Butler}. 
Let us also mention a recent state-of-the-art quadrupole-octupole 
symmetry-projected configuration mixing study for $^{144}$Ba 
\cite{Tomas_GCM_parity_2016}.
 
Given the experimental interest in studying octupole properties of nuclei
heavier than Th, we consider in the present work the dynamical interplay 
between quadrupole  and octupole degrees of freedom 
in a selected set of  even-even actinides, i.e., $^{220-240}$U, $^{222-242}$Pu, $^{222-242}$Cm
and $^{222-242}$Cf. These nuclei have $Z$ values away from $Z=88$ (Ra) which is
considered to be a ``magic number" for the existence of permanent octupole
deformation \cite{butler_96}. 
The study of the dynamical quadrupole-octupole
coupling in the selected actinide nuclei
allows us to examine the role of the corresponding zero-point quantum fluctuations 
on the systematic of  the $1^{-}$ excitation energies, transition strengths and  
correlation energies 
around 
the  $N = 134$ (a neutron octupole magic number) isotones $^{226}$U, $^{228}$Pu, 
$^{230}$Cm and $^{232}$Cf. 

As in our previous study \cite{Rayner_Q2Q3_GCM_2012}, we consider three 
levels of approximation for each of the studied nuclei. The constrained 
Gogny-HFB scheme is used to obtain MFPESs as functions of both the 
quadrupole and octupole moments. As discussed later, those MFPESs  can 
be rather soft along the octupole direction. Some of the considered 
nuclei also exhibit transitional features along the quadrupole 
direction. In this case the HFB approximation can only be considered as 
a starting point and  beyond-mean-field correlations should be taken 
into account.  First, parity projection is carried out in order to 
build the corresponding parity-projected potential energy surfaces 
(PPPESs). Next, both symmetry restoration as well as  fluctuations in 
the collective quadrupole and octupole coordinates are taken into 
account within the 2D-GCM framework. Although reflection symmetry is 
also restored by our GCM ansatz (see, Sec.~\ref{GCM-Theory-used}), the 
parity-projected results allow us to disentangle the relative 
contribution to the total correlation that has to be associated with 
the restoration of the reflection symmetry.

All the results discussed in this 
paper have been obtained with the Gogny-D1M EDF \cite{gogny-d1m}. Among the members of the D1 family of  
parametrizations of the Gogny-EDF, D1S \cite{gogny-d1s} 
has already built a strong reputation among practitioners, given its ability to reproduce 
a wealth of low-energy nuclear data all over the nuclear chart  both at the mean-field level and beyond
(see, for example, Ref.~\cite{Review_RoToRa_2019} and references therein). Nevertheless, the parametrization 
D1M, specially tailored to better describe nuclear masses, has already provided 
a reasonable description of nuclear properties
in different regions of the nuclear 
chart (see, for example, Refs.~\cite{ours-PT,ours-Y-Nb-quasi,Rayner-fission-1,Rayner-fission-5}
and references therein). In particular, previous
studies \cite{JPG_2012_RoRay,Robledo-Bertsch-Q3-1,Rayner_Q2Q3_GCM_2012,Robledo_2D-GCM_with_Butler}
have shown that the parametrization D1M essentially keeps the same predictive power as D1S when applied to 
the description of octupole properties.

\begin{figure}
\includegraphics[width=0.47\textwidth]{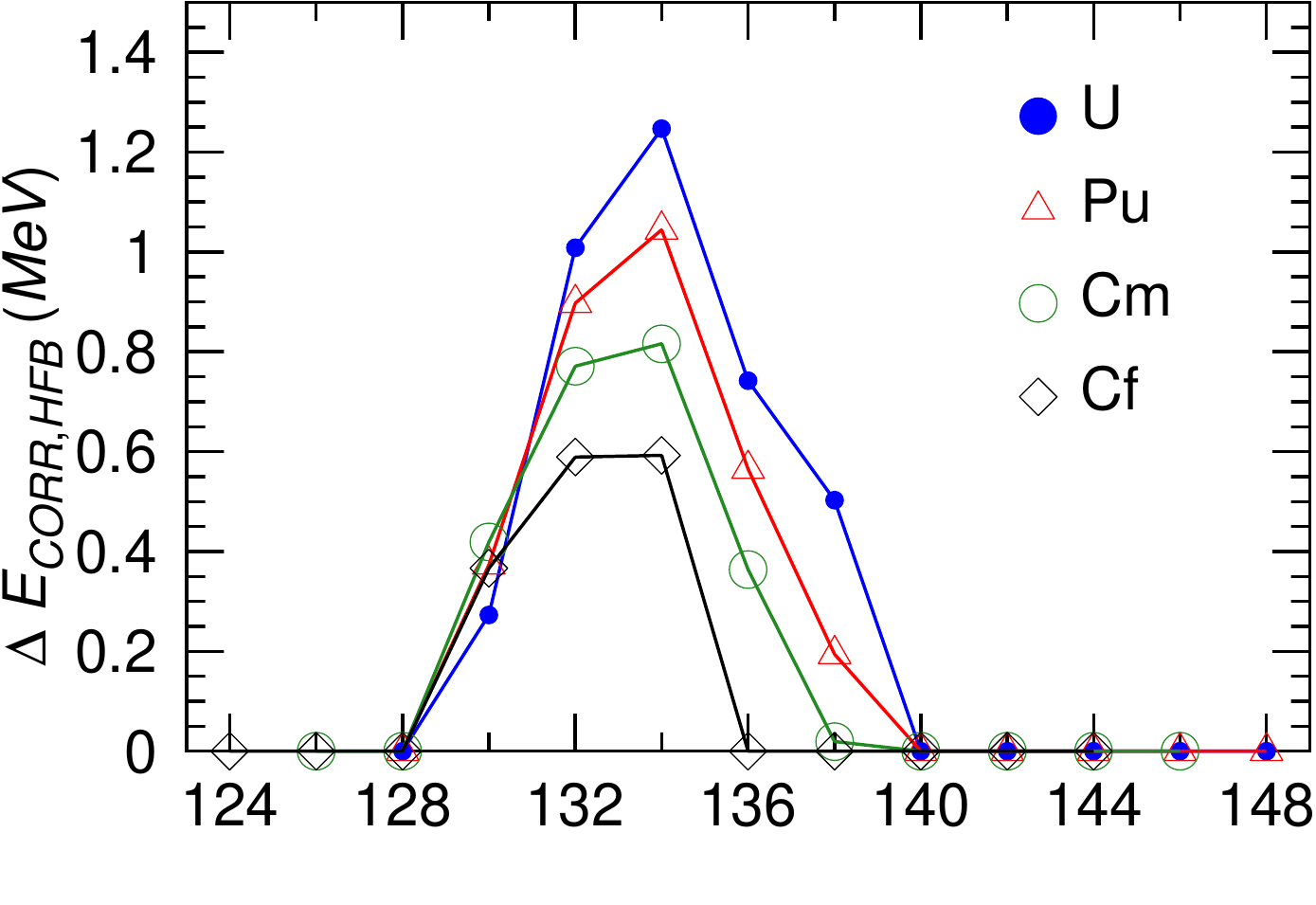}
\caption{(Color online) The mean-field octupole correlation 
energies Eq.(\ref{MFCorrEner}) are plotted as functions
of the neutron number. Results have been obtained with
the Gogny-D1M EDF. For more details, see the main text. 
}
\label{MF_Summary} 
\end{figure}

The paper is organized as follows. The different approaches employed in 
this work are briefly outlined in Secs.~\ref{MF-Theory-used}, 
\ref{Parity-Theory-used} and \ref{GCM-Theory-used}. In each section the 
results obtained with the corresponding approaches are discussed. 
Mean-field results are presented in Sec.~\ref{MF-Theory-used}. We then 
turn our attention to beyond-mean-field properties, i.e., parity 
restoration and configuration mixing in Secs.~\ref{Parity-Theory-used} 
and \ref{GCM-Theory-used}. Special attention is paid in 
Sec.~\ref{GCM-Theory-used} to $1^{-}$ energy splittings, reduced 
transition probabilities, correlation energies and their comparison 
with the available experimental data \cite{EXP-DATA}. Finally, 
Sec.~\ref{conclusions} is devoted to the concluding remarks.

\section{Results}

The aim of this work is to study  the quadrupole-octupole 
dynamics in a selected set of actinide  nuclei. Three
levels of approximation have been considered: the HFB approach \cite{rs}
with 
constrains on the (axially symmetric) quadrupole and octupole 
operators,  parity projection 
and the 2D-GCM. In what follows, we outline those 
approaches \cite{Rayner_Q2Q3_GCM_2012,Robledo_2D-GCM_with_Butler}, based on the Gogny-D1M EDF, and
discuss the results obtained with each of them.

\subsection{Mean-field}
\label{MF-Theory-used}

\begin{figure*}
\includegraphics[width=0.9\textwidth]{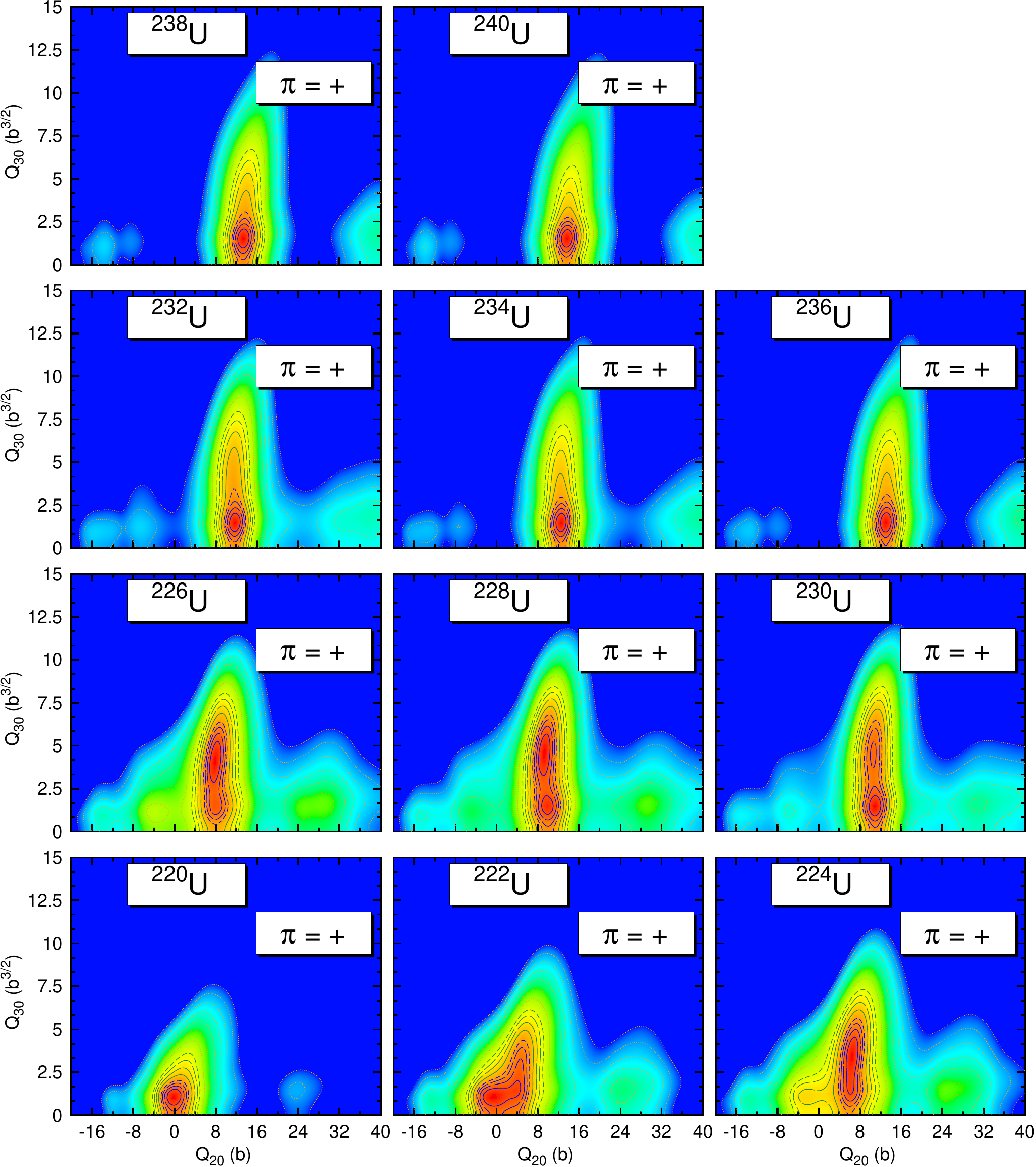}
\caption{(Color online) Positive $\pi = +1$ parity-projected potential energy surfaces (PPPESs)
computed with the Gogny-D1M EDF for the isotopes $^{220-240}$U. See, caption of 
Fig.~\ref{Q2Q3_MF_U} for the contour-line patterns.
}
\label{Q2Q3_POS_PARITY_U} 
\end{figure*}

To obtain the MFPESs,  the HFB equation with constrains on the 
axially symmetric quadrupole  
\begin{equation}
\hat{Q}_{20} = z^{2} - \frac{1}{2} \Big(x^{2} + y^{2} \Big)
\end{equation}
and octupole operator
\begin{equation}
\hat{Q}_{30} = z^{3} - \frac{3}{2} \Big(x^{2} + y^{2} \Big)z
\end{equation} 
is solved. The mean value with the HFB intrinsic state $| \Phi \rangle$
of the two operators define the quadrupole and octupole deformation parameters
$Q_{20}$ and $Q_{30}$. From them one can compute \cite{egi92} the standard deformation parameters
$\beta_l=\sqrt{4\pi(2l+1)}/(3R_0^l A) Q_{l0}$ with $R_0=1.2A^{1/3}$ \footnote{For $A=230$
a value of $Q_{20}=1000$ fm$^2$ is equivalent to $\beta_2=0.212$ and a value of
$Q_{30}=1000$ fm$^3$ is equivalent to $\beta_3=0.034$.} In order
to alleviate the already substantial 
computational effort, both axial and time-reversal
symmetries have been kept as self-consistent symmetries. The HFB equation 
is solved using a performing, approximate second-order gradient method \cite{rob11}.
The center of mass is fixed at the origin to avoid spurious 
effects associated with its motion 
\cite{Rayner_Q2Q3_GCM_2012,Robledo_2D-GCM_with_Butler}. The HFB quasiparticle
operators \cite{rs} have been expanded in a deformed (axially symmetric) harmonic 
oscillator (HO) basis containing 16 major shells to grant convergence for the studied
physical quantities.

\begin{figure*}
\includegraphics[width=0.9\textwidth]{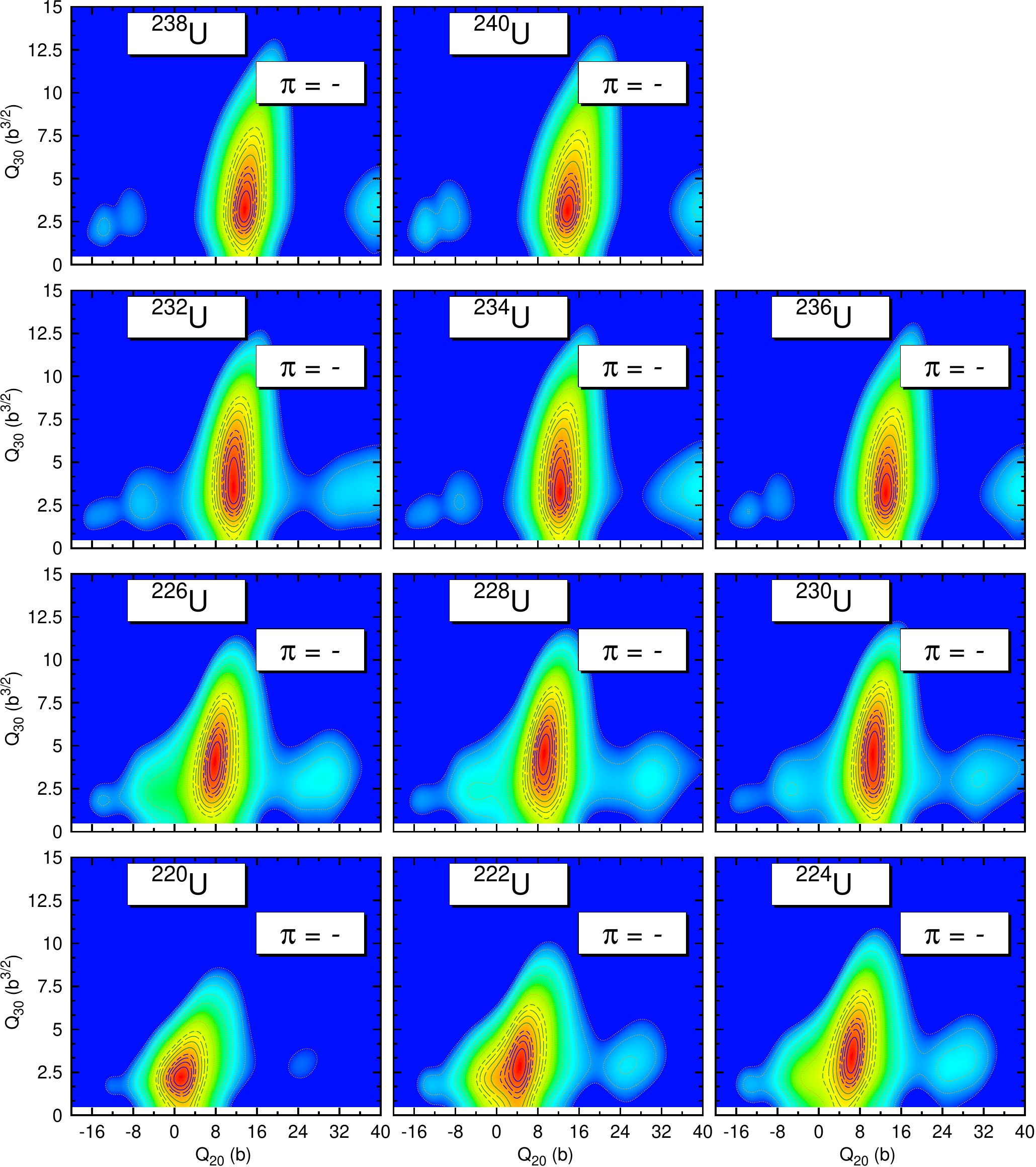}
\caption{(Color online) Negative $\pi = -1$ parity-projected potential energy surfaces (PPPESs)
computed with the Gogny-D1M EDF for the isotopes $^{220-240}$U. See, caption of 
Fig.~\ref{Q2Q3_MF_U} for the contour-line patterns.
}
\label{Q2Q3_NEG_PARITY_U} 
\end{figure*}

The $(Q_{20},Q_{30})$-constrained Gogny-HFB calculations 
provide a set of states $| \Phi ({\bf{Q}})\rangle$ labeled by their corresponding 
static deformations ${\bf{Q}} =(Q_{20},Q_{30})$. The HFB energies $E_{HFB}({\bf{Q}})$ associated
with those states define the contour plots referred to as  
MFPESs in this work. As the Gogny-EDF is invariant under 
parity transformation \cite{rod02,egi04} the associated HFB energies satisfy the property
$E_{HFB}(Q_{20},Q_{30})= E_{HFB}(Q_{20},-Q_{30})$. For this reason, only positive 
octupole moments are considered when plotting  PESs.

The MFPESs obtained for the  isotopes $^{220-240}$U  are 
shown in Fig.~\ref{Q2Q3_MF_U}  as illustrative examples.
In our calculations, the  $Q_{20}$-grid   
$-20 \textrm{b} \le Q_{20} \le 40 \textrm{b}$ (with a step $\delta Q_{20} = 1 \textrm{b} $) 
and the $Q_{30}$-grid  $0 \textrm{b}^{3/2} \le Q_{30} \le 15 \textrm{b}^{3/2}$
(with a step $\delta Q_{30} = 0.5  \textrm{b}^{3/2}$) have been employed. 
Along the $Q_{20}$-direction there is a shape/phase  transition from a
spherical  ground state in $^{220}$U to a well quadrupole deformed ground state in $^{240}$U.
A similar structural  evolution along the $Q_{20}$-direction 
have been 
obtained for the Pu, Cm, and Cf isotopic chains. Spherical
or weakly deformed ground states are obtained for isotopes with $N \approx 126$ 
while a well quadrupole deformed ground state emerges with increasing neutron number.
In fact, we have obtained (static) HFB quadrupole deformations 
within the range  $0 \textrm{b} \le Q_{20,GS} \le 14 \textrm{b}$. The only exception is the nucleus 
$^{222}$Cf for which $Q_{20,GS} = -2 $~b. Many of the considered 
isotopes  exhibit octupole deformation in their HFB ground state: 
$^{222-230}$U, $^{224-232}$Pu, $^{226-234}$Cm and 
$^{228-232}$Cf with values of the octupole moment in 
the range $ 2 \textrm{b}^{3/2} \le Q_{30,GS} \le 5 \textrm{b}^{3/2}$. 

\begin{figure}
\includegraphics[width=0.42\textwidth]{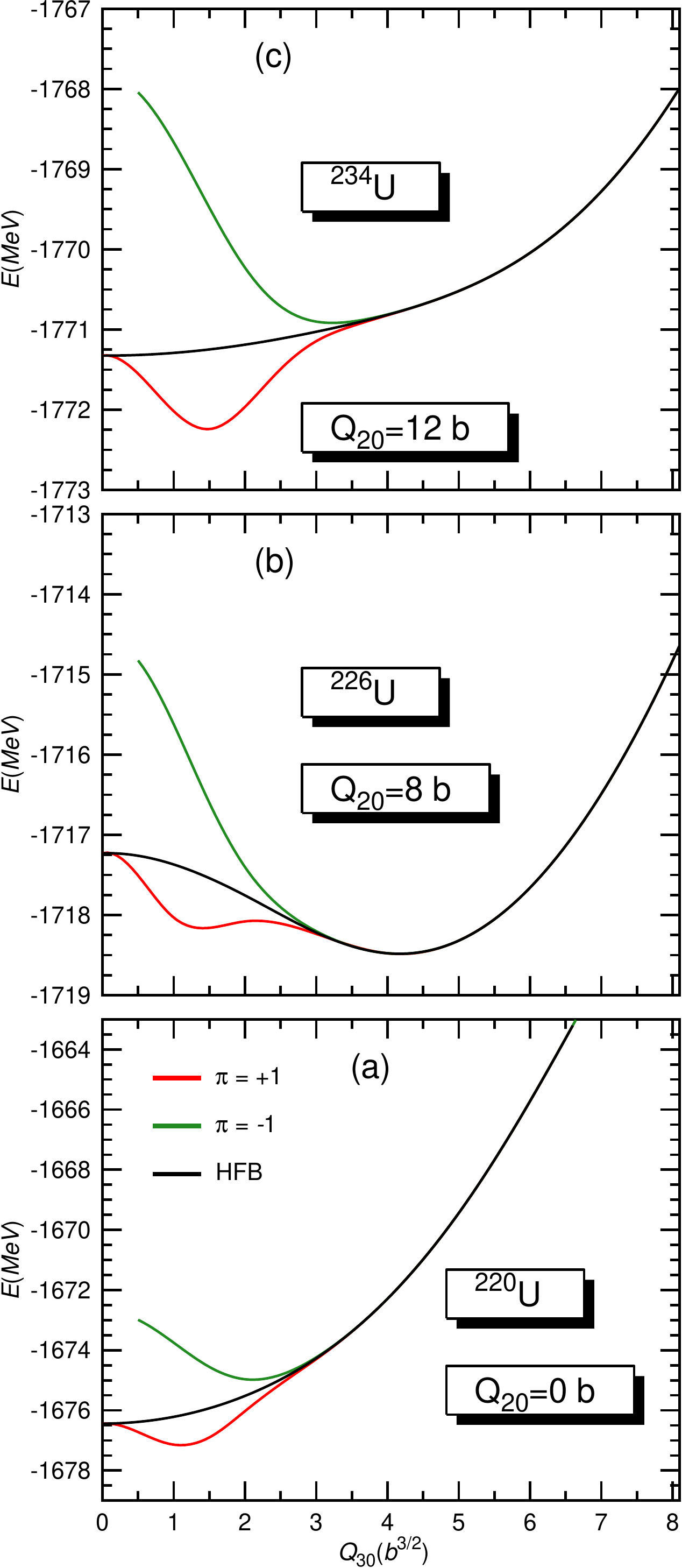}
\caption{(Color online) The $\pi = +1$ (red) and $\pi = -1$ (green) parity-projected energies
are depicted as functions of the octupole moment $Q_{30}$ for fixed values of the quadrupole moment
$Q_{20}$ in the nuclei $^{220}$U, $^{226}$U and $^{234}$U.
The corresponding HFB energies are also included in the plots. Results have been obtained 
with the Gogny-D1M EDF.
}
\label{FOLLOW-Fig} 
\end{figure}

The MFPESs depicted in 
Fig.~\ref{Q2Q3_MF_U}, as well as the ones obtained for the Pu, Cm and Cf
isotopic chains are rather soft along the $Q_{30}$-direction. This 
is further illustrated in Fig.~\ref{FOLLOW-Fig} where
the HFB energies obtained for 
$^{220}$U, $^{226}$U and $^{234}$U have been plotted, as functions of $Q_{30}$, for 
fixed values of the 
quadrupole moment corresponding  to the  absolute minima of the PESs.

The mean-field octupole correlation energies defined as
the energy gained by allowing octupolarity in the ground state 
\begin{equation}
\label{MFCorrEner}
\Delta E_{CORR, HFB} = E_{HFB, Q_{30}=0} - E_{HFB,GS}
\end{equation}
are plotted in Fig.~\ref{MF_Summary}. The largest values
($1.25$, $1.04$, $0.81$ and $0.59$ $MeV$) correspond
to  $N = 134$ isotones. Note, that the 
relatively small energies $E_{CORR, HFB}$ result from 
the softness observed in the MFPESs of 
nuclei with octupole deformed 
ground states [see, for example, panel (b) of Fig.~\ref{FOLLOW-Fig}].

\begin{figure}
\includegraphics[width=0.47\textwidth]{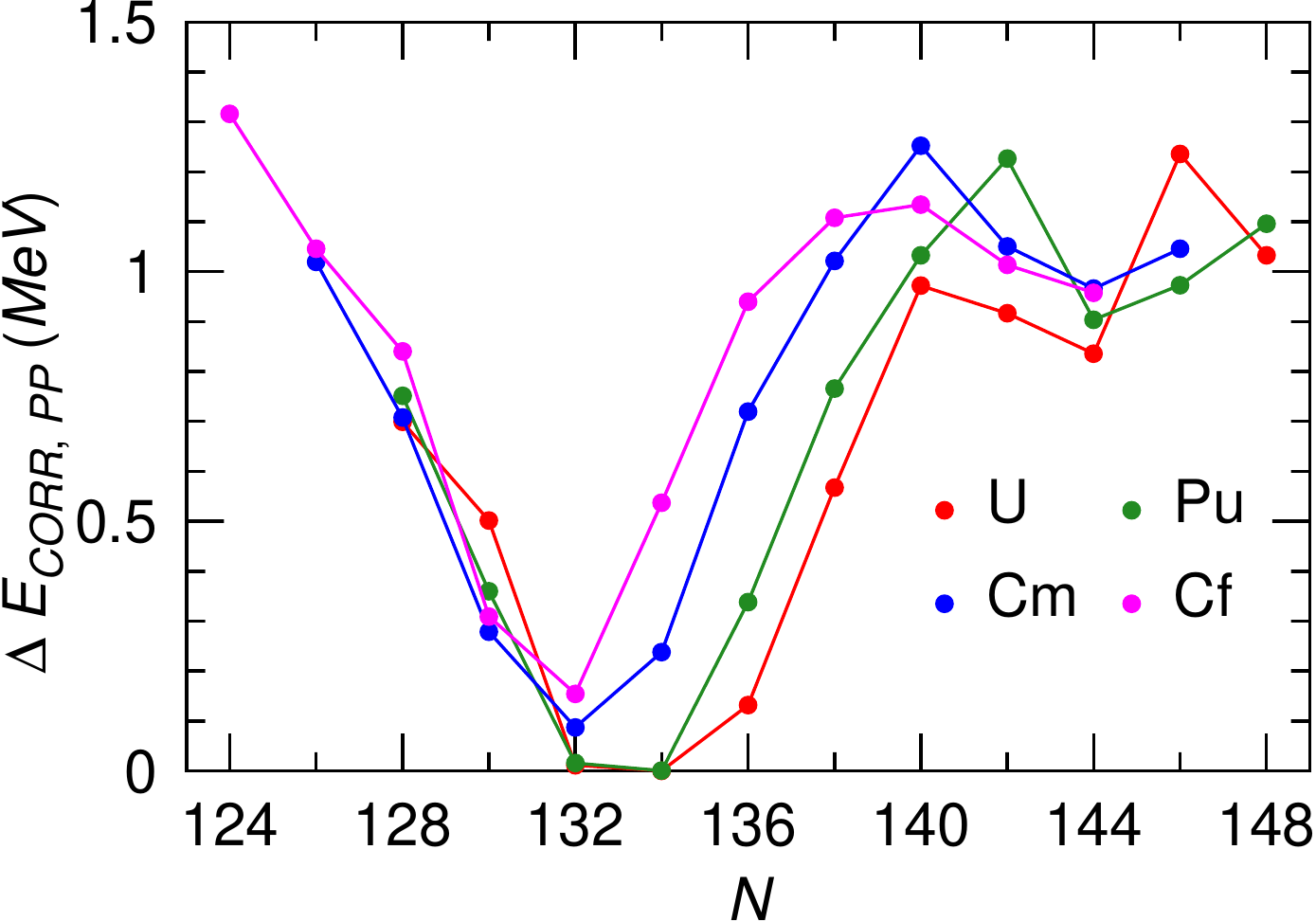}
\caption{(Color online) The correlation energies
stemming from parity restoration 
Eq.(\ref{PPCorreEner})
are plotted as functions of the neutron number. Results have been obtained 
with the Gogny-D1M EDF. For more details, see the main text.
}
\label{PPCorrEner-Fig} 
\end{figure}

The softness of the MFPESs discussed in this section already point towards the
key role of dynamical beyond-mean-field correlations, i.e., symmetry 
restoration and/or quadrupole-octupole configuration mixing in the studied nuclei. 
Two spatial symmetries are broken in 
 this study. One is the rotational symmetry with the quadrupole moment as the 
 relevant parameter and the other is the reflection symmetry with the octupole 
 moment as the relevant parameter. From the previous discussion of mean-field 
 results it is clear that the octupole is the softest mode. Therefore, parity is the
 most important symmetry to be restored. It would be desirable to restore also
 both the rotational and particle number symmetries. This kind of simultaneous symmetry 
 restoration is feasible in lighter nuclear systems. However, when combined with the quadrupole-octupole
 configuration mixing of Sec.~\ref{GCM-Theory-used}, it becomes a highly 
 demanding computational task  \cite{Tomas_GCM_parity_2016} 
 out of the scope of an exhaustive survey like the one discussed in this paper.

\begin{figure*}
\includegraphics[width=0.9\textwidth,angle=-90]{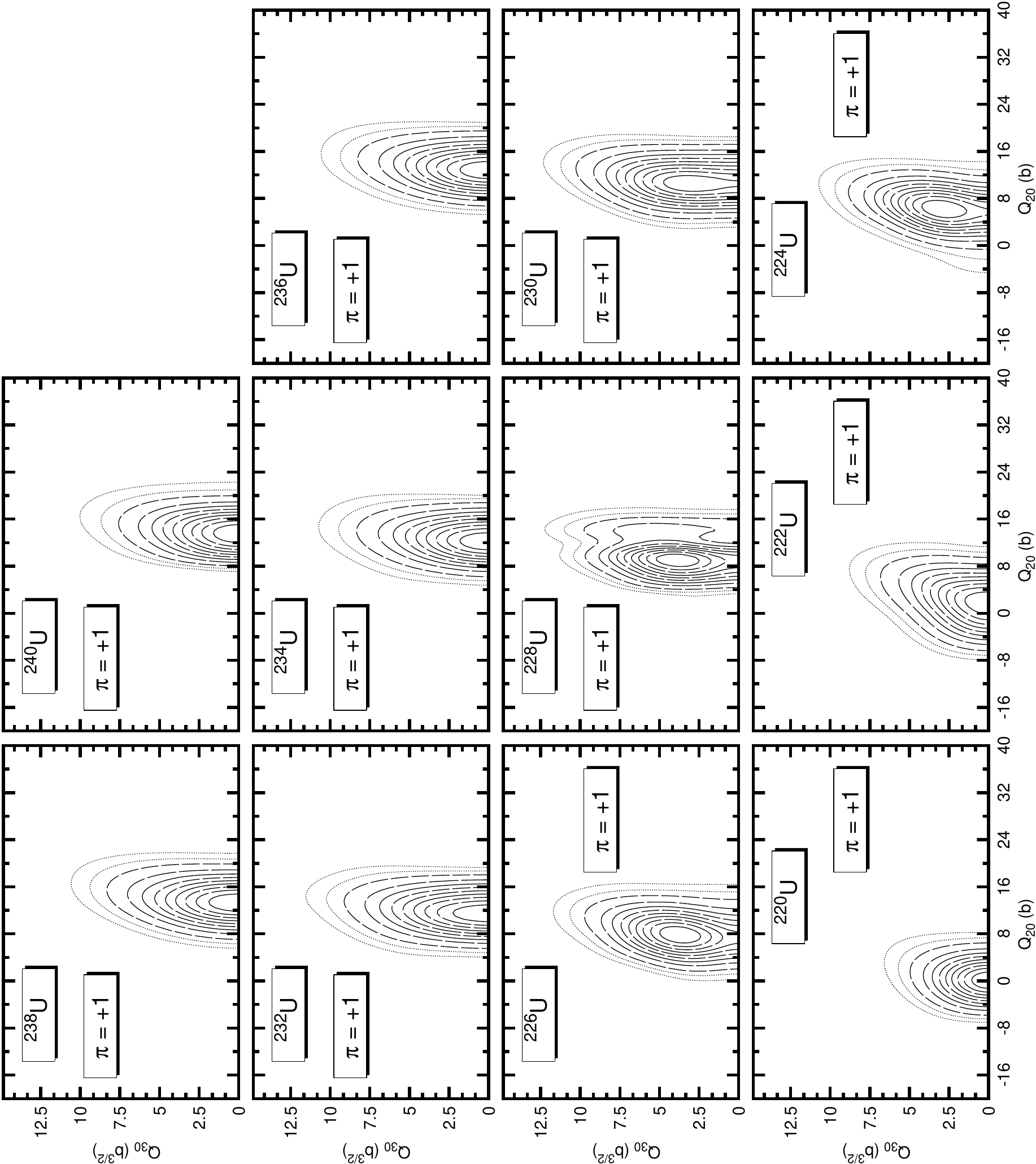}
\caption{Collective wave functions Eq.(\ref{cll-wfs-HW}) squared for the ground states of the nuclei
$^{220-240}$U. The contour lines (a succession of solid, long dashed and short dashed lines) start 
at $90 \%$ of the maximum value up $10 \%$ of it. The two dotted-line contours correspond to the 
tail of the amplitude ($15 \%$ and $1 \%$ of the maximum value). Results have been obtained with the 
Gogny-D1M EDF. For more details, see the main text.
}
\label{COLLWS_POS_PARITY_U} 
\end{figure*}


\subsection{Parity symmetry restoration}
\label{Parity-Theory-used}

Parity symmetry is broken by intrinsic HFB states with a non-zero value of
the octupole moment. To  restore the symmetry \cite{mar83,egi91,Rayner_Q2Q3_GCM_2012}
we build parity-projected states $| \Phi^{\pi} ({\bf{Q}})\rangle$ from 
the intrinsic HFB states $| \Phi ({\bf{Q}})\rangle$ by acting on them with the parity projector 
\begin{equation}
\hat{{\cal{P}}}^{\pi} = \frac{1}{2} \left(1 + \pi \hat{\Pi} \right),
\end{equation}
where $\pi=\pm 1$ is the desired parity quantum number. For each of the projected
states with parity $\pi$ one can compute the projected energy
\begin{eqnarray} \label{PROJEDF}
E_{\pi} ({\bf Q}) &=&
\frac{
\langle {\Phi} ({\bf Q}) | \hat{H} [\rho(\vec{r})] | {\Phi} ({\bf Q}) \rangle
}
{
\langle {\Phi} ({\bf Q}) |  {\Phi} ({\bf Q}) \rangle
+
\pi \langle {\Phi} ({\bf Q}) |  \hat{\Pi} | {\Phi} ({\bf Q}) \rangle
}
\nonumber\\
\nonumber\\
&+& \pi
\frac{\langle {\Phi} ({\bf Q}) |
\hat{H} [\theta(\vec{r})]  \hat{\Pi} | {\Phi} ({\bf Q}) \rangle
}
{
\langle {\Phi} ({\bf Q}) |  {\Phi} ({\bf Q}) \rangle
+
\pi \langle {\Phi} ({\bf Q}) |  \hat{\Pi} | {\Phi} ({\bf Q}) \rangle
}
\end{eqnarray}

\begin{figure*}
\includegraphics[width=0.9\textwidth,angle=-90]{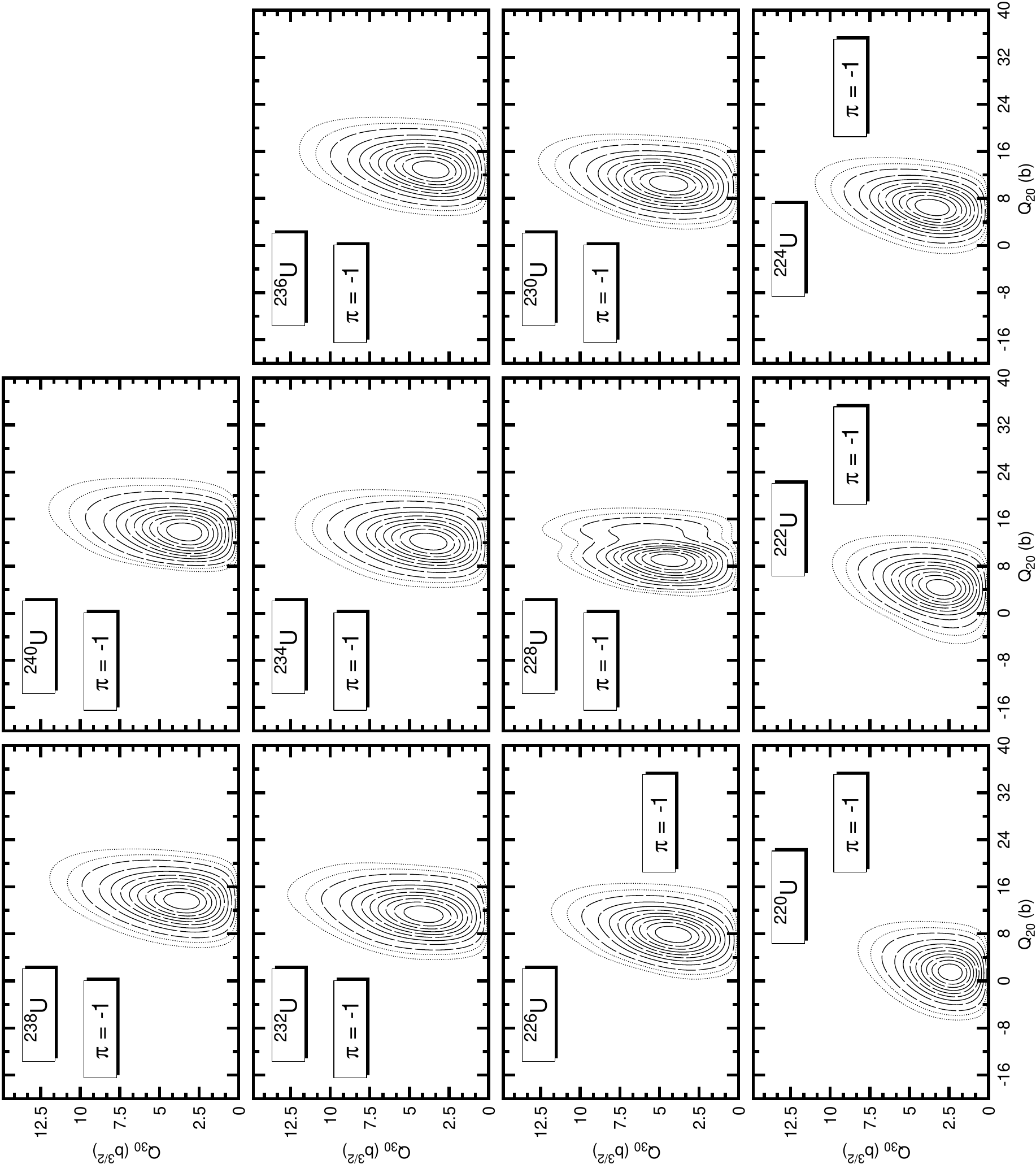}
\caption{Collective wave functions Eq.(\ref{cll-wfs-HW}) squared for the 
lowest negative-parity states of the nuclei
$^{220-240}$U. See, caption of Fig.~\ref{COLLWS_POS_PARITY_U} for contour-line patterns. 
Results have been obtained with the 
Gogny-D1M EDF. For more details, see the main text. 
}
\label{COLLWS_NEG_PARITY_U} 
\end{figure*}

The evaluation of the Hamiltonian overlaps 
$\langle {\Phi} ({\bf Q}) | \hat{H} [\rho(\vec{r})] | {\Phi} ({\bf Q}) \rangle$
and 
$\langle {\Phi} ({\bf Q}) |
\hat{H} [\theta(\vec{r})]  \hat{\Pi} | {\Phi} ({\bf Q}) \rangle$  
in Eq.(\ref{PROJEDF}) requires a prescription for the 
density-dependent part of the Gogny-EDF. As in previous 
studies \cite{Rayner_Q2Q3_GCM_2012,Robledo_2D-GCM_with_Butler}, we 
use the {\it mixed density} prescription that amounts
to consider the  densities
\begin{equation}
\rho(\vec{r})= 
\frac{
\langle {\Phi} ({\bf Q}) | \hat{\rho}({\vec{r})} | {\Phi} ({\bf Q}) \rangle
}
{
\langle {\Phi} ({\bf Q}) | {\Phi} ({\bf Q}) \rangle
},
\end{equation}
and 
\begin{equation}
\theta(\vec{r})= 
\frac{
\langle {\Phi} ({\bf Q}) | \hat{\rho}({\vec{r})} \hat{\Pi} | {\Phi} ({\bf Q}) \rangle
}
{
\langle {\Phi} ({\bf Q}) | \hat{\Pi} | {\Phi} ({\bf Q}) \rangle
}
\end{equation}
Such a prescription  guarantees various consistency requirements within the EDF
framework 
and avoids pathologies associated with the restoration of spatial 
symmetries \cite{rod02,egi04,robledo_presciption-1,robledo_presciption-2}. The parity-projected  proton 
and neutron numbers, 
usually differ from the nucleus' proton  $Z_{0}$ and neutron $N_{0}$ numbers. To correct the energy
for this deviation we have 
replaced   $\hat{H}$ by 
$\hat{H} -  \lambda_{Z} \left( \hat{Z} -Z_{0} \right)-  \lambda_{N} \left( \hat{N} -N_{0} \right)$, where 
$\lambda_{Z}$ and $\lambda_{N}$ are chemical potentials 
for protons and neutrons, respectively \cite{har82,bon90,Rayner_Q2Q3_GCM_2012,Robledo_2D-GCM_with_Butler}.

The $\pi =+1$ and $\pi =-1$ PPPESs obtained for the  isotopes $^{220-240}$U  are 
depicted in Figs.~\ref{Q2Q3_POS_PARITY_U} and \ref{Q2Q3_NEG_PARITY_U} as illustrative examples.
Along the $Q_{30}=0$ axis, the projection onto positive parity is unnecessary as the 
corresponding quadrupole deformed even-even intrinsic states are already 
pure $\pi =+1$ states. On the other hand, in the case of negative parity, the evaluation 
of the projected energy along the $Q_{30}=0$ axis requires to resolve a "zero-over-zero" 
indeterminacy \cite{egi91,Rayner_Q2Q3_GCM_2012}. However, the $\pi = -1$ 
projected energy increases rapidly when approaching $Q_{30}=0$ (see, Fig.~\ref{FOLLOW-Fig}) and its
limiting value does 
not play a significant role in the discussion of the PPPESs. We have then omitted 
this quantity along the $Q_{30}=0$ axis
in Fig.~\ref{Q2Q3_NEG_PARITY_U}.

The absolute minima of the $\pi =+1$ and $\pi =-1$ PPPESs are located at  quadrupole deformations 
close to the  HFB values discussed in Sec.~\ref{MF-Theory-used}. In the case of the 
$\pi =+1$ PPPESs, depicted in Fig.~\ref{Q2Q3_POS_PARITY_U},  a characteristic pocket 
develops
with a minimum at $Q_{30} = 1.0-1.5 b^{3/2}$.
In the case 
of nuclei with a reflection-symmetric  
HFB ground state, such a minimum is 
the global one. This is illustrated in panels (a) and (c) of  Fig.~\ref{FOLLOW-Fig}
where the $\pi =+1$ parity-projected energies 
obtained for $^{220}$U and $^{234}$U
are plotted, as functions of $Q_{30}$,  for fixed values of the quadrupole moment 
corresponding  to the absolute minima of the PESs.
On the other hand, for nuclei with a reflection-asymmetric 
mean-field ground state, there is a pronounced competition with a second minimum at
$Q_{30} = 3.0-4.5 b^{3/2}$ as illustrated in panel (b) of 
Fig.~\ref{FOLLOW-Fig} for $^{226}$U. In the case of $^{226}$U, the 
global $\pi =+1$  minimum  at $Q_{30} = 4.0 b^{3/2}$ is only $500 KeV$ deeper than the one at 
$Q_{30} = 1.0 b^{3/2}$.   Similar results have been obtained for Pu, Cm and Cf isotopes.
For example, the global $\pi =+1$ minima correspond to $Q_{30} = 3.0 b^{3/2}$ and $4.0 b^{3/2}$ in 
$^{226}$~Pu and $^{228}$Pu, respectively, while for other  Pu isotopes 
as well as for  Cm and Cf nuclei they are located at 
$Q_{30} = 1.0-1.5 b^{3/2}$. 
As can be seen from Figs.~\ref{Q2Q3_MF_U}, \ref{Q2Q3_POS_PARITY_U} and 
\ref{FOLLOW-Fig} not only the MFPESs but also the $\pi =+1$ PPPESs are rather 
soft along the $Q_{30}$-direction.

The $\pi =-1$ PPPESs, depicted in Fig.~\ref{Q2Q3_NEG_PARITY_U}, display
well developed absolute minima at $Q_{30}= 2.0-4.5 b^{3/2}$. In the case 
of nuclei with a reflection-symmetric  
HFB ground state, such as $^{220}$U and $^{234}$U, the absolute $\pi =-1$ minima have larger octupole deformations 
than the $\pi =+1$ ones [see, panels (a) and (c) of  Fig.~\ref{FOLLOW-Fig}]. On the other hand, for some nuclei
with a reflection-asymmetric HFB ground state, such as $^{226}$U, the (almost degenerate)
$\pi =-1$  and $\pi =+1$ absolute 
minima  have similar octupole deformations [see, panel (b) of 
Fig.~\ref{FOLLOW-Fig}]. Similar features have been found for the other isotopic chains. Let us 
mention, that the complex topography
along the $Q_{30}$-direction as well as the  transition 
to an octupole-deformed regime found in our Gogny-D1M 
calculations has also been studied, as a function of the 
strength of the two-body interaction, in Ref.~\cite{LMG-model} using
the parity-projected Lipkin-Meshkov-Glick (LMG) model.

\begin{figure}
\includegraphics[width=0.47\textwidth]{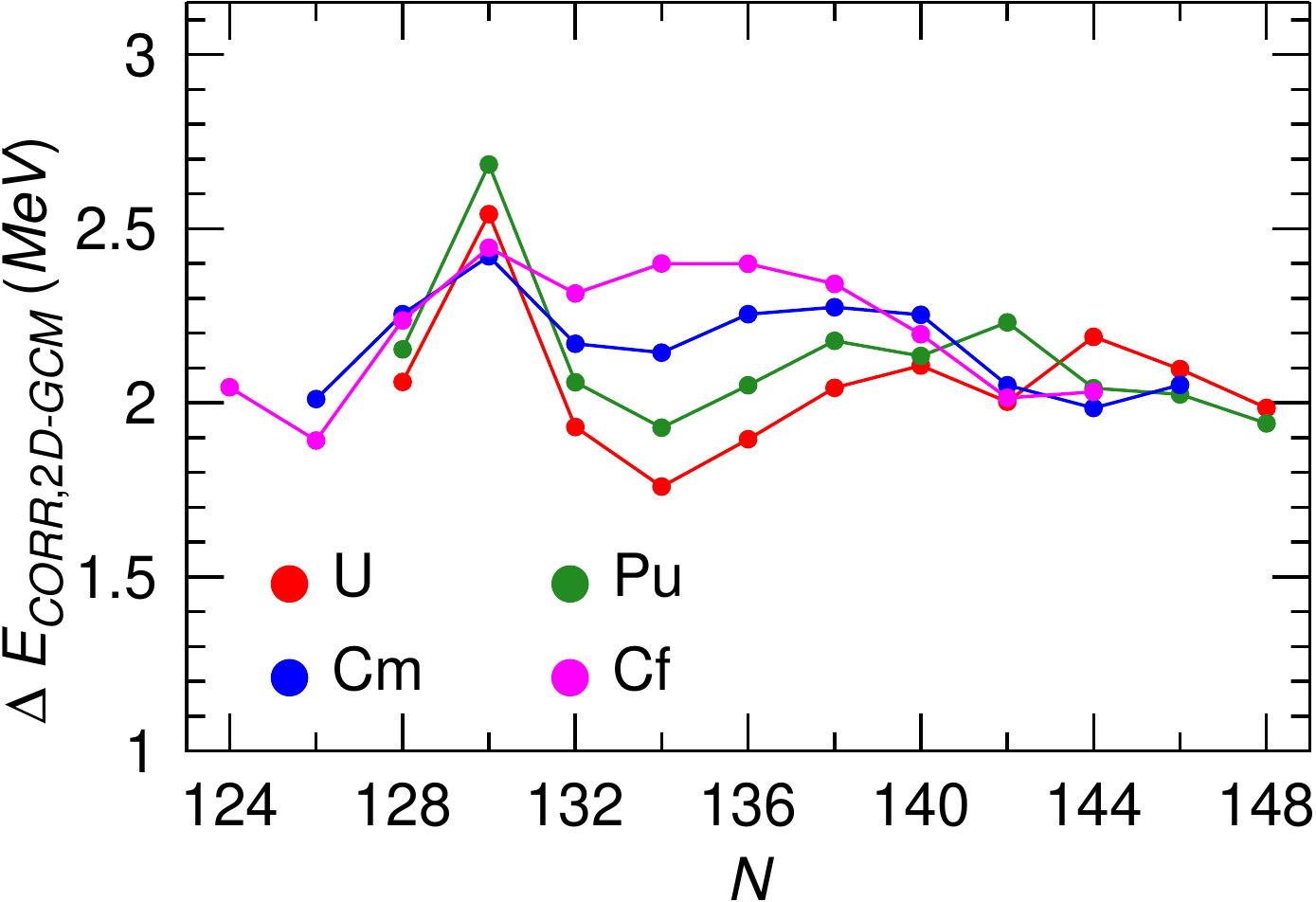}
\caption{(Color online) 
The correlation energies
obtained within the 2D-GCM framework 
Eq.(\ref{2DGCMCorreEner})
are plotted as functions of the neutron number. Results have been obtained 
with the Gogny-D1M EDF. For more details, see the main text.
}
\label{ECORR-2D-GCM} 
\end{figure}

As a measure of the correlations induced by parity symmetry restoration one can use the
correlation energy, defined in terms of the difference between the 
HFB  $E_{HFB, GS}$ and parity projected $E_{\pi=+1,GS}$  ground state energies
\begin{equation}
\label{PPCorreEner}
\Delta E_{CORR, PP} = E_{HFB, GS} - E_{\pi=+1,GS}.
\end{equation}
In Fig.~\ref{PPCorrEner-Fig}, we show this quantity for the different isotopes 
considered. The correlation energy shows a minimum around $N = 132-134$
corresponding to strongly octupole-deformed intrinsic states.
As  shown later on in Sec.~\ref{GCM-Theory-used}, the comparison 
between the correlation energies $E_{CORR, PP}$ and the ones obtained within the 
symmetry-conserving 2D-GCM framework 
(see, Fig.~\ref{ECORR-2D-GCM})
reveals the key role
played by quantum fluctuations around those neutron numbers.

\subsection{Generator Coordinate Method}
\label{GCM-Theory-used}


%

We include quantum fluctuations in the quadrupole and octupole degrees 
of freedom by considering a linear superposition of the HFB states $| \Phi 
({\bf{Q}})\rangle$
\begin{equation} \label{GCM-WF}
| {\Psi}_{\sigma}^{\pi} \rangle = \int d{\bf Q} f_{\sigma}^{\pi} ({\bf Q}) | {\Phi} ({\bf Q}) \rangle
\end{equation}
where, both positive and negative 
octupole moments  $Q_{30}$ are included in the integration domain. In this
way the parity of the collective amplitude under the change of sign of $Q_3$, 
namely $f_{\sigma}^{\pi}(Q_{20},-Q_{30})=\pi f_{\sigma}^{\pi}(Q_{20},Q_{30})$, 
determines the parity of $|\Psi_\sigma^\pi\rangle$.
The property $f_{\sigma}^{\pi}(Q_{20},-Q_{30})=\pi f_{\sigma}^{\pi}(Q_{20},Q_{30})$ 
is a direct consequence of the 
invariance of the interaction under the parity symmetry operation.  
The index $\sigma$ in Eq.(\ref{GCM-WF}) labels the different GCM solutions. 

The amplitudes $f_{\sigma}^{\pi} ({\bf Q})$ are solutions of the 
Griffin-Hill-Wheeler (GHW) equation \cite{rs}
\begin{equation} \label{HW-equation}
\int d{\bf Q}^{'} 
\left(
{\cal{H}}({\bf Q},{\bf Q}^{'}) - E_{\sigma}^{\pi}
{\cal{N}}({\bf Q}, {\bf Q}^{'})
\right)
f_{\sigma}^{\pi} ({\bf Q}^{'}) = 0.
\end{equation}
with the Hamiltonian and norm kernels defined in the standard way 
\begin{eqnarray} \label{GCM-PROJEDF-hnk}
{\cal{H}}({\bf Q}, {\bf Q}^{'}) &=&
\langle {\Phi} ({\bf Q}) | 
\hat{H} [\rho^{GCM}(\vec{r}) ] | {\Phi} ({\bf Q}^{'}) \rangle,
\nonumber\\
{\cal{N}}({\bf Q}, {\bf Q}^{'}) &=&
\langle {\Phi} ({\bf Q}) |  {\Phi} ({\bf Q}^{'}) \rangle
\end{eqnarray}

In the evaluation of the Hamiltonian kernel ${\cal{H}}({\bf Q}, {\bf Q}^{'})$
for the  Gogny-EDF, we 
have employed the {\it mixed density} prescription 
\begin{equation}
\rho^{GCM}(\vec{r})= 
\frac{
\langle {\Phi} ({\bf Q}) | \hat{\rho}({\vec{r})} | {\Phi} ({\bf Q}^{'}) \rangle
}
{
\langle {\Phi} ({\bf Q}) | {\Phi} ({\bf Q}^{'}) \rangle
}.
\end{equation}
As in the parity projection case, first-order corrections to take into account
deviations in both the proton and neutron numbers
\cite{har82,bon90,Rayner_Q2Q3_GCM_2012,Robledo_2D-GCM_with_Butler} are included.

The HFB basis states $| {\Phi} ({\bf Q}) \rangle$ are not
orthonormal. Therefore, the amplitudes $f_{\sigma}^{\pi} ({\bf Q})$ 
cannot be interpreted as probability amplitudes. 
Instead, one considers the so-called collective wave functions 
\begin{equation} \label{cll-wfs-HW} 
G_{\sigma}^{\pi} ({\bf Q}) =   \int d{\bf Q}^{'} {\cal
{N}}^{\frac{1}{2}}({\bf Q}, {\bf Q}^{'})  f_{\sigma}^{\pi}({\bf 
Q}^{'}),
\end{equation}  
written in terms of the square root operator 
${\cal
{N}}^{\frac{1}{2}}({\bf Q}, {\bf Q}^{'})$ of the norm kernel \cite{rs,rod02,Rayner_Q2Q3_GCM_2012} 
defined by the property
\begin{equation} \label{Oversqrt}
{\cal{N}}({\bf Q}; {\bf Q}^{'})  = \int d{\bf Q}^{''} 
{\cal{N}}^{\frac{1}{2}}({\bf Q}; {\bf Q}^{''}) {\cal{N}}^{\frac{1}{2}}({\bf Q}^{''}; {\bf Q}^{'})
\end{equation}

The overlap  $ \langle \Psi_{\sigma}^{\pi} | \hat{O} |\Psi_{\sigma '}^{\pi '} \rangle$
of an operator $\hat{O}$ between two different GCM states 
Eq.(\ref{GCM-WF})
is required in the computation of physical quantities such as, for example, the 
electromagnetic transition probabilities. It reads 
\begin{eqnarray} \label{OOverGCM}
\langle \Psi_{\sigma}^{\pi} | \hat{O} |\Psi_{\sigma '}^{\pi '} \rangle =
\int d {\bf Q} d {\bf Q}^{'} G_{\sigma}^{\pi \,*} ({\bf Q}) {\cal O } ({\bf Q}, {\bf Q}^{'})
G_{\sigma '}^{\pi '} ({\bf Q}^{'})
\end{eqnarray}
where
\begin{eqnarray} \label{Okernel}
{\cal{O}} ({\bf Q}, {\bf Q}^{'})  &=& \int d{\bf Q}^{''} d{\bf Q}^{'''}
{\cal{N}}^{-\frac{1}{2}}({\bf Q}; {\bf Q}^{''}) \langle {\bf Q}^{''} | \hat{O}
| {\bf Q}^{'''} \rangle  \times
\nonumber\\
& \times &
{\cal{N}}^{-\frac{1}{2}}({\bf Q}^{'''}; {\bf Q}^{'})
\end{eqnarray}

For the reduced transition probabilities 
$B(E1,1^{-} \rightarrow 0^{+} )$ and 
$B(E3,3^{-} \rightarrow 0^{+} )$ the rotational 
formula for K=0 bands have been used 
\begin{equation} \label{ROT_FORMULA}
	B(E\lambda,\lambda^{-} \rightarrow 0^{+} ) = \frac{e^{2}}{4 \pi} 
	\Big{|} \langle \Psi_{\sigma}^{\pi=-1} | \hat{{\cal O}}_{\lambda} 
	|\Psi_{{\sigma}^{'}=1}^{{\pi}^{'}=+1} \rangle \Big{|}^{2}.
\end{equation}
For $B(E1)$ and $B(E3)$ transitions $\sigma$ corresponds to the first excited GCM state with negative 
parity. The electromagnetic transition operators $\hat{{\cal O}}_{1}$
and $\hat{{\cal O}}_{3}$ are the dipole moment operator and 
the proton component of the octupole operator, respectively \cite{Rayner_Q2Q3_GCM_2012}.

Some comments are in order here regarding the use of 
Eq.(\ref{ROT_FORMULA}). Previous 
studies \cite{B_Ro_limitations_ROT,limitations_ROT_Rob_EPJA} have  revealed 
that the use of proper angular momentum projected (AMP) wave functions concurs in 
an enhancement of the $E3$  strengths in spherical and/or weakly 
quadrupole deformed nuclei as compared to 
the strength obtained with the rotational formula implicit in Eq.(\ref{ROT_FORMULA}). 
On the other hand, the 
$E1$ transitions do not show a clear pattern due to their less
collective nature. With this in mind, the 
$E3$ strengths obtained in our calculations for 
spherical and/or weakly deformed 
$N \approx 126$ nuclei via Eq.(\ref{ROT_FORMULA}), should be 
viewed as lower bounds.

The collective wave functions Eq.(\ref{cll-wfs-HW}) squared 
corresponding to the ground and lowest negative parity 2D-GCM 
states in $^{220-240}$U
are plotted  
in Figs.~\ref{COLLWS_POS_PARITY_U} 
and \ref{COLLWS_NEG_PARITY_U}, respectively. As can be seen from
Fig.~\ref{COLLWS_POS_PARITY_U},  
the ground state collective amplitudes
$|G_{\sigma=1}^{\pi=+1} (Q_{20},Q_{30}) |^{2}$ reach global maxima 
for octupole moments  different from 
zero only in $^{224-230}$U. The same holds for  
$^{226-232}$Pu and  $^{228,230}$Cm while for other U, Pu and Cf
nuclei, the peaks are located around $Q_{30} =0$. 
As illustrated in Fig.~\ref{COLLWS_POS_PARITY_U}, the spreading of 
the amplitudes 
$|G_{\sigma=1}^{\pi=+1} (Q_{20},Q_{30}) |^{2}$
along the $Q_{30}$-direction is large, indicating the 
octupole-soft character of the $\pi =+1$
2D-GCM ground states. In the case of the 
$\pi=-1$  amplitudes, depicted in Fig.~\ref{COLLWS_NEG_PARITY_U}, the 
maxima are always located at a nonzero octupole moment  as could be 
anticipated from the behavior of the $\pi=-1$ PPPESs (see, Fig.~\ref{Q2Q3_NEG_PARITY_U}).

Using Eq.(\ref{OOverGCM}), we have computed the 2D-GCM average 
quadrupole moments 
\begin{equation}
	(\bar{Q}_{20})_{\sigma}^{\pi}= 
	\langle {\Psi}_{\sigma}^{\pi} | \hat{Q}_{20}| {\Psi}_{\sigma}^{\pi} \rangle.
\end{equation}
In the case of a negative-parity operator like $\hat{Q}_{30}$  the quantity
$\langle {\Psi}_{\sigma}^{\pi} | {\hat{Q}}_{30}| {\Psi}_{\sigma}^{\pi} \rangle$
is zero by construction. Therefore, a meaningful  
averaged quantity has to be defined \cite{Rayner_Q2Q3_GCM_2012} by restricting
the integration domain $\mathcal{D}$ to positive values of $Q_{30}$ and $Q_{30}^{'}$ 
\begin{equation}
	(\bar{Q}_{30})_{\sigma}^{\pi}= 4 \int_{\mathcal{D}} d {\bf Q} d {\bf Q'} G_{\sigma}^{\pi \,*} ( {\bf Q} ) 
	 {\cal Q }_{30} ({\bf Q}, {\bf Q}^{'})
G_{\sigma}^{\pi} ({\bf Q}^{'})
\end{equation}
In the case of a strongly peaked collective inertia, the average octupole moment $\bar{Q}_{30}$
is a good estimator of the location of the peak.

The ground-state dynamical quadrupole moments  $(\bar{Q}_{20})_{\sigma=1}^{\pi=+1}$ 
increase as more neutrons are added along a given isotopic chain and their values 
remain close to the ones predicted at the HFB level. On the other hand, at variance
with the HFB results, once both $\pi =+1$ symmetry restoration and 
$(Q_{20},Q_{30})$-fluctuations are considered at the 2D-GCM level, dynamical octupole 
deformations $0.53 b^{3/2} \le (\bar{Q}_{30})_{\sigma=1}^{\pi=+1} \le 2.15 b^{3/2}$
are found in the ground states of all the studied nuclei with  the largest
values corresponding  to isotopes with  neutron numbers $N = 132-138$. 
The quadrupole moments $(\bar{Q}_{20})_{\sigma}^{\pi=-1}$ 
corresponding to the lowest negative-parity states 
also increase 
their values with increasing $N$. Moreover, the corresponding 
average octupole moments lie within the range
$1.87 b^{3/2} \le (\bar{Q}_{30})_{\sigma}^{\pi=-1} \le 3.75 b^{3/2}$ with 
their largest values being reached once more  for  $N = 132-138$ isotopes.

\begin{figure*}
\includegraphics[width=0.9\textwidth]{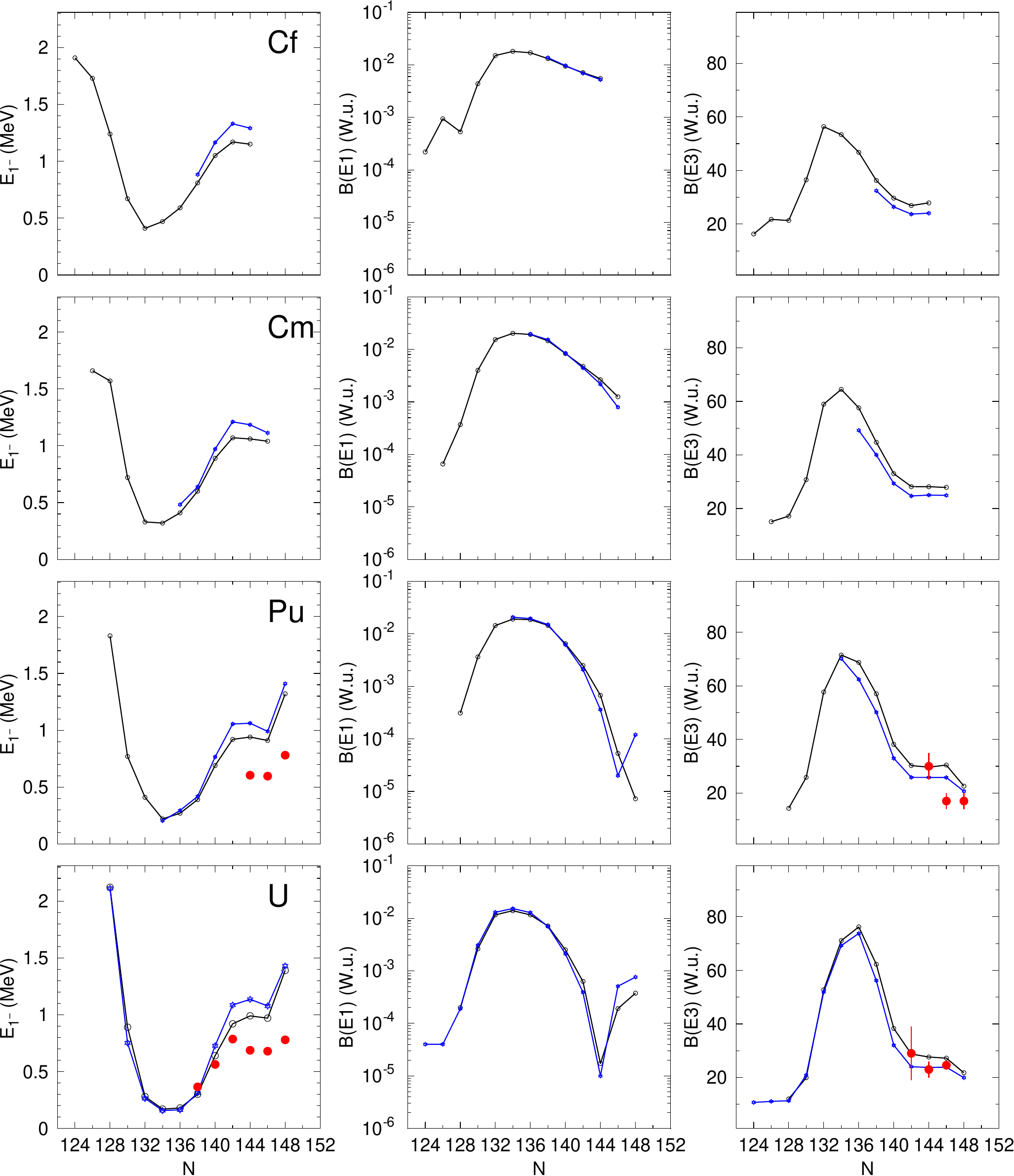}
\caption{(Color online) The 2D-GCM $E_{1^{-}}$ energy splittings (left panels) and the  reduced transition 
probabilities 
$B(E1)$ (middle panels) and $B(E3)$ (right panels) are plotted 
(in black) as functions of the neutron number for the studied U, Pu, Cm and Cf 
isotopic chains.  The available experimental data (in red) have been taken from Ref.~\cite{EXP-DATA}.
The $E_{1^{-}}$, $B(E1)$ and $B(E3)$ values obtained in the framework of the 
1D-GCM \cite{Robledo-Bertsch-Q3-1}, with
the octupole moment as single  generating coordinate, have also been included 
(in blue)
in each of the plots.
Results have been obtained 
with the Gogny-D1M EDF. For more details, see the main text.
}
\label{BE1BE3E1-summary} 
\end{figure*}

The correlation energies, defined as the difference between the  
HFB and 2D-GCM ground-state energies
\begin{equation}
\label{2DGCMCorreEner}
\Delta E_{CORR, 2D-GCM} = E_{HFB, GS} - E_{\pi=+1,2D-GCM}
\end{equation}
are depicted in Fig.~\ref{ECORR-2D-GCM}. They exhibit a weaker dependence 
with  neutron number than the $\Delta E_{CORR, PP}$ values
stemming from parity restoration (see, Fig.~\ref{PPCorrEner-Fig}). 
The inclusion of beyond-mean-field correlations, via the 2D-GCM ansatz 
Eq.(\ref{GCM-WF}), substantially modifies the behavior 
observed in  Fig.~\ref{PPCorrEner-Fig} around the neutron numbers $N = 132-134$ providing  
a smoother trend. Furthermore, the  variation of 
the correlation energies (within the range $1.76 \,\textrm{MeV} \le \Delta E_{CORR, 2D-GCM} \le 2.46 \,\textrm{MeV}$)
is of the same order of magnitude as the rms for the binding energy in Gogny-like nuclear mass 
tables \cite{gogny-d1m} and, therefore, those correlation energies should be 
considered in improved versions of the Gogny-EDF.

The energy difference $E_{1^{-}}$ between the positive parity ground state and the lowest 
$1^-$ excited state, obtained in the 2D-GCM calculations, is shown in 
the left panels of 
Fig.~\ref{BE1BE3E1-summary} as a function of the neutron number. The energies are  very small for 
$^{224-230}$U in agreement with their large (dynamical) octupole deformation. 
Other U isotopes, with less pronounced  dynamical  octupole 
deformation effects, display larger $E_{1^{-}}$ values and the first negative parity excited state
can be interpreted as an 
octupole vibrational state. In the same panels, we have also included 
the  energy differences $E_{1^{-}}$ obtained within the 
framework of the 1D-GCM with the octupole moment as single generating 
coordinate \cite{Robledo-Bertsch-Q3-1}. As can be seen, the trend with neutron number is similar 
in both  calculations. However, for heavier isotopes the 
2D-GCM $E_{1^{-}}$ energies tend to be  smaller than the 1D-GCM ones. Regarding 
the comparison with the the available experimental data, we are able to 
reproduce the increase of the excitation energies with increasing 
neutron number. However, exception made of the $N= 138-140$ isotopes, 
the predicted $E_{1^{-}}$ energies are larger than the experimental ones, a feature 
found in many GCM calculations 
(see, for example, \cite{Robledo-Bertsch-Q3-1,Rayner_Q2Q3_GCM_2012}). Similar 
results are found for the other isotopic chains.

In the case of the $B(E1)$  reduced transition probabilities, depicted 
in the middle panels of Fig.~\ref{BE1BE3E1-summary}, no experimental data are available. 
Exception made of the nucleus $^{242}$Pu, the 1D-GCM and 2D-GCM calculations display a 
similar
pattern with 
the largest $B(E1)$ values corresponding  to
the neutron numbers $N=132-136$. As discussed in \cite{egi90} ,the $B(E1)$ strength 
strongly depend on how the dipole moment evolves with octupole deformation in the region
where the positive and negative parity wave functions overlap. In the $^{242}$Pu
case the dipole moment changes sign in the region of interest and there is 
a strong cancellation depending upon subtle details of the collective wave
functions. For other nuclei, however, the sign of the dipole moment does
not change with octupole deformation and the dependency with the details
of the collective wave functions is much weaker. Although the 1D and 2D
GCM collective wave functions look very similar, the tiny differences can
easily explain the differences in the results of the two calculations. 
Note, that both approaches 
predict a pronounced minimum for $^{236}$U also consequence of a dipole
moment changing its sign as the octupole moment increases.
The  $B(E3)$ reduced transition probabilities 
are plotted in the right panels of Fig.~\ref{BE1BE3E1-summary}. They show marked 
maxima for $N = 132-136$ that correlate well with the features observed for
the  $E_{1^{-}}$ energies and the $B(E1)$  strengths. Though  
essentially the same trend is obtained, for heavier nuclei the 2D-GCM 
$B(E3)$ values are larger than the 1D-GCM ones. As can be seen from
the panels, the predicted $B(E3)$
strengths for $^{234-238}$U and $^{238-242}$Pu compare reasonably well
with the available experimental data.

Finally, let us mention 
that the comparison between the 2D-GCM and 1D-GCM results in 
Fig.~\ref{BE1BE3E1-summary} reveals that, to a large extent, there is
a decoupling between the quadrupole and octupole degrees of freedom 
in the studied nuclei and confirms that the 1D-GCM approach 
\cite{Robledo-Bertsch-Q3-1} represents
a valuable computational tool to account for the systematic of the $1^{-}$ energy splittings
and reduced transition probabilities in this region of the nuclear chart.

\begin{figure*}
\includegraphics[width=0.3\textwidth,angle=-90]{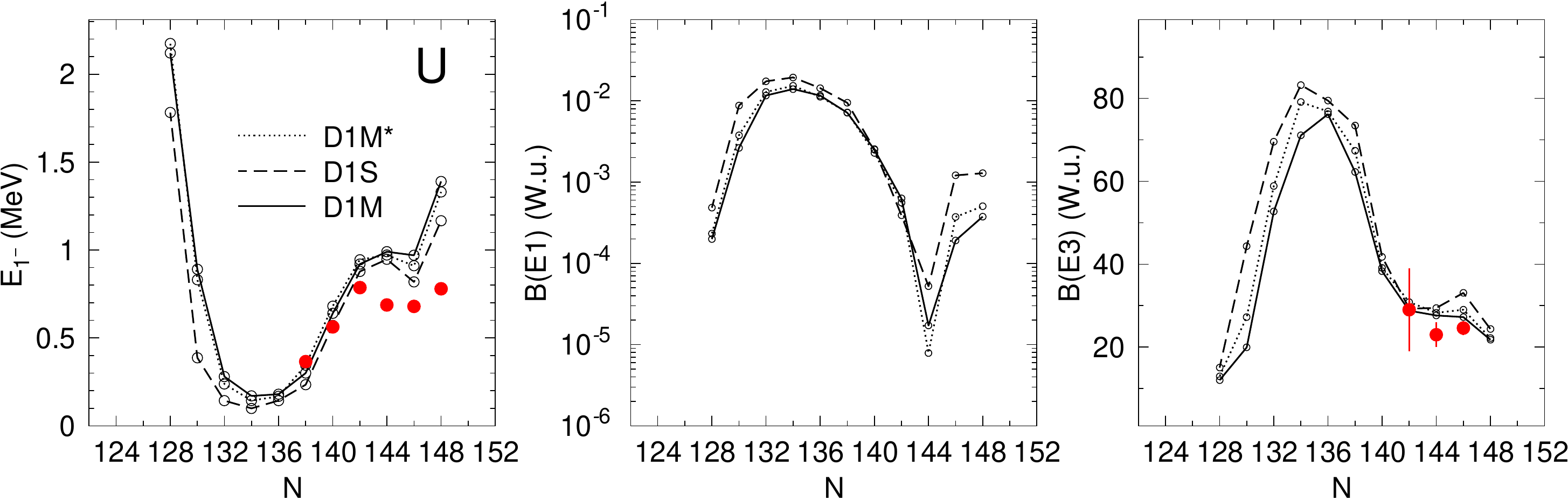}
\caption{(Color online) Same as Fig \ref{BE1BE3E1-summary} but for different
parametrizations of the Gogny force (D1M full line, D1M* dotted line and D1S dashed line).
}
\label{Fig11} 
\end{figure*}

In order to explore the robustness of the results with a change of the parametrization
of the interaction, we have carried out in the uranium chain the same kind of 2D GCM calculations
but with the D1S and D1M* parametrizations of the Gogny force. The later is a
newly proposed re-parametrization of D1M with the goal of improving the slope
of the symmetry energy \cite{gonzalez18} while preserving as much as possible
other properties of D1M. The results are shown in Fig \ref{Fig11} along with
the experimental data.

The trend with neutron number is similar in the three calculations confirming
the consistency of the results. There 
are some quantitative differences at $N=130$ where a transition from octupole
soft to octupole deformed ground state takes place. Those differences are larger for 
D1S as expected, because D1M* was fitted to be as close as possible to D1M.
From the comparison we conclude that the trend of the results with neutron
number is rather insensitive to the interaction used.

\section{Conclusions}
\label{conclusions}

In this paper we have studied the interplay between the quadrupole and octupole degrees 
of freedom in a selected set of even-even actinides  both at the mean-field level and 
beyond. To this end, we have resorted to the static Gogny-HFB approach, parity projection 
as well as 2D-GCM calculations with the multipole moments $Q_{20}$ and $Q_{30}$ as generating 
coordinates. At the mean-field level only nuclei with 
neutron numbers $ 130 \le N \le 138$  exhibit octupole deformed HFB ground states. However, for 
all the studied nuclei, the  MFPESs and PPPESs are rather soft along the 
$Q_{30}$-direction. As a result, once 
correlations associated with parity restoration and quadrupole-octupole configuration
mixing are included simultaneously within the 2D-GCM approach, their ground states 
turn out to be (dynamically) octupole deformed, albeit with the largest
octupole deformation effects still corresponding to $N=132-138$ isotopes. Moreover, within the
2D-GCM approach, the correlation energies display  a weaker dependence on the neutron 
number. Given the range of variation of those 2D-GCM correlation energies, they should 
be included in the fitting protocol of improved versions of the Gogny-EDF. Using the 
correlated 2D-GCM states, we have studied the systematic of the $1^{-}$  
energy splittings as well as $B(E1)$ and $B(E3)$ reduced transition probabilities in the 
considered isotopic chains. The predicted values compare reasonably well with
the available experimental data. They  point towards a (dynamically) 
enhanced octupolarity for $N=132-138$ isotopes while octupole-vibrational
states have been found for other nuclei. The comparison with 1D-GCM results 
\cite{Robledo-Bertsch-Q3-1}
reveals
that, for the studied nuclei, the quadrupole-octupole
coupling is  weak and to a large extent the  properties of negative 
parity states (i.e., energy splittings and 
reduced transition probabilities) can be
reasonably well described in terms of the octupole degree of freedom alone.

\begin{acknowledgments}
The  work of LMR was supported by Spanish Ministry 
of Economy and Competitiveness (MINECO) Grants No. 
PGC2018-094583-B-I00.
\end{acknowledgments}


\begin{thebibliography}{00}

\bibitem{Ahmad_93} I. Ahmad and 
P. A. Butler, Ann. Rev. Nucl. Part. Sci. {\bf 43}, 71 (1993).

\bibitem{butler_2016} P. A. Butler, J. Phys. G {\bf 43}, 073002 (2016).

\bibitem{butler_2015} P. A. Butler and L. Willmann, Nucl. Phys. News {\bf 25}, 12 (2015).


\bibitem{butler_96} P.A. Butler and W. Nazarewicz, Rev. Mod. Phys.
{\bf 68}, 349 (1996).


\bibitem{Tandel_2013} S. K. Tandel, M. Hemalatha, A. Y. Deo, S. B. Patel, R. Palit, 
T. Trivedi, J. Sethi, S. Saha, D. C. Biswas 
and S. Mukhopadhyay, Phys. Rev. C {\bf 87}, 034319 (2013).

\bibitem{Gaffney_2013} L. P. Gaffney {\it et al}., Nature {\bf 497}, 199 (2013).


\bibitem{Li_2014} H. J. Li, S. J. Zhu, J. H. Hamilton, E. H. Wang, A. V. Ramayya,
Y. J. Chen, J. K. Hwang, J. Ranger, S. H. Liu, Z. G. Xiao, Y. Huang, Z. Zhang, Y. X. Luo,
J. O. Rasmussen, I. Y. Lee, G. M. Ter-Akopian, Y. T. Oganessian  
and W. C. Ma, Phys. Rev. C {\bf 90}, 047303 (2014).

\bibitem{Ahmad_2015} I. Ahmad, R. R. Chasman, J. P. Green, F. G. Kondev 
and S. Zhu, Phys. Rev. C {\bf 92}, 024313 (2015).

\bibitem{Bucher_2016} B. Bucher {\it et al}., Phys. Rev. Lett. {\bf 116}, 112503 (2016).

\bibitem{Bucher_146Ba_2017} B. Bucher {\it et al}, Phys. Rev. Lett. {\bf 118},  152504  (2017).

\bibitem{Butler2020} P. A. Butler, L. P. Gaffney, {\it et al}, 
                     Phys. Rev. Lett. \textbf{124}, 042503 (2020).

\bibitem{Chishti20} M. M. R. Chishti, D. O’Donnell, {\it et al}
                   Nature Physics (2020)

\bibitem{moller_81} P. M\"oller and J.R. Nix, Nucl. Phys. {\bf A361}, 117 (1981).

\bibitem{leander_82} G.A. Leander, R.K. Sheline, P. M\"oller, P. Olanders,
                  I. Ragnarsson, and A.J. Sierk,  
                  Nucl. Phys. {\bf A388}, 452 (1982).

\bibitem{naza_84} W. Nazarewicz {\em et al.}, 
                  Nucl. Phys. {\bf A429}, 269 (1984).

\bibitem{naza_92} W. Nazarewicz and S.L. Tabor, 
                  Phys. Rev. C {\bf 45}, 2226 (1992).
		  
\bibitem{babilon_05} M. Babilon, N.V. Zamfir, D. Kusnezov, E.A. McCutchan,
and A. Zilges, Phys. Rev. C {\bf 72}, 064302 (2005).
		  
\bibitem{minkov_06} N. Minkov, P. Yotov, S. Drenska, W. Scheid, D. Bonatsos,
D. Lenis, and D. Petrellis, Phys. Rev. C {\bf 73}, 044315 (2006).
		  

\bibitem{Bing_2014} K. Nomura, D. Vretenar, T. Niksic and  Bing-Nan Lu, , Phys. Rev. C {\bf 89}, 024312 (2014).

\bibitem{Nomura_Ba_RE_2018} K. Nomura, T. Niksic and D. Vretenar, Phys. Rev. C {\bf 97}, 024317 (2018).

\bibitem{Nomura_Th_RE_2018} K. Nomura, D. Vretenar and  B. -N. Lu, Phys. Rev. C {\bf 88}, 021303 (2013). 

\bibitem{Nomura_Rayner_IBM_2015} K. Nomura, R. Rodr\'iguez-Guzm\'an 
and L. M. Robledo, Phys. Rev. C {\bf 92}, 014312 (2015).


\bibitem{mar83}  S. Marcos, H. Flocard, and P.H. Heenen,
                 Nucl. Phys. {\bf A410}, 125 (1983).

\bibitem{bon86}  P. Bonche, P. -H. Heenen, H. Flocard, and D. Vautherin,
                 Phys. Lett. {\bf B175}, 387 (1986).

\bibitem{bon91}  P. Bonche, J.S. Krieger, M.S. Weiss, J. Dobaczewski, 
                 H. Flocard, and P.-H. Heenen, 
                 Phys. Rev. Lett. {\bf 66}, 876 (1991).

\bibitem{hee94}  P.-H. Heenen, J. Skalski, P.Bonche, and H. Flocard, 
                 Phys. Rev. C {\bf 50},802 (1994). 

\bibitem{erler-85}  J. Erler, K. Langanke, H. P. Loens, G. Mart\'inez-Pinedo 
and  P.-G.  Reinhard, Phys. Rev. C {\bf 85}, 025802 (2012). 

\bibitem{Ebataba-2017} S. Ebata and T. Nakatsukasa,  Physica Scr. {\bf 92}, 064005 (2017).



\bibitem{rob87}  L.M. Robledo, J.L. Egido, J.F. Berger, and M. Girod,
                 Phys. Lett. {\bf B187}, 223 (1987).

\bibitem{rob88}  L.M. Robledo, J.L. Egido, B. Nerlo-Pomorska, and K. Pomorski,
                 Phys. Lett. {\bf B201}, 409 (1988).

\bibitem{egi90}  J.L. Egido and L.M. Robledo, 
                 Nucl. Phys. {\bf A518}, 475 (1990).

\bibitem{egi91}  J.L. Egido and L.M. Robledo, 
                 Nucl. Phys. {\bf A524}, 65 (1991).

\bibitem{gar98}  E. Garrote, J.L. Egido, and L.M. Robledo, 
                 Phys. Rev. Lett. {\bf 80}, 4398 (1998); 
                 Nucl. Phys. {\bf A654}, 723c (1999).

\bibitem{rob10}  L.M. Robledo, M. Baldo, P. Schuck, and X. Vi\~nas, 
                 Phys. Rev. C {\bf 81},034315 (2010).

\bibitem{egi92}  J.L. Egido and L.M. Robledo, 
                 Nucl. Phys. {\bf A545}, 589 (1992).

\bibitem{Fission-D1Mstarstar} R. Rodr\'iguez-Guzm\'an, Y. M. Humadi 
and L. M. Robledo, Eur. Phys. J. A, {\bf 56}, 43 (2020).		   


\bibitem{Long_2004} W. H. Long, J. Meng, N. Van Giai and S. G. Zhou, 
                   Phys. Rev. C {\bf 69}, 034319 (2004).
		   		   

\bibitem{Xia_PRC_2017} S. Y. Xia, H. Tao, Y. Lu, Z. P. Li, T. Niksic 
and D. Vretenar, Phys. Rev. C {\bf 96}, 054303  (2017).

\bibitem{Xu-2017}  Z. Xu  and  Z.-P. Li, Chinese Phys. C {\bf 41}, 124107 (2017).


\bibitem{Agbemava-Q3-2016}  S. E. Agbemava, A. V. Afanasjev and P. Ring, Phys. Rev. C {\bf 93}, 044304 (2016).

\bibitem{Agbemava-Q3-2017} S. E. Agbemava and  A. V. Afanasjev, Phys. Rev. C {\bf 96}, 024301 (2017).

\bibitem{Recent-Survey-Q3} Y. Cao, S. E. Agbemava, A. V. Afanasjev, W. Nazarewicz 
and E. Olsen, ArXiv:2004.01319v1  [nucl-th].

\bibitem{Tomas_GCM_parity_2016} R\'emi N. Bernard, Luis M. Robledo 
and Tom\'as R. Rodr\'iguez, Phys. Rev C  {\bf 93},  061302 (2016).


\bibitem{JPG_2012_RoRay} L. M. Robledo and 
R. Rodr\'iguez-Guzm\'an, J. Phys. G: Nucl. Part. Phys. {\bf 39}, 105103 (2012). 


\bibitem{gogny-d1s} 
J. F. Berger, M. Girod, and D. Gogny, 
Nucl. Phys. A {\bf{428}}, 23c (1984).

\bibitem{gogny-d1n} 
F. Chappert, M. Girod, and S. Hilaire, 
Phys. Lett. B {\bf{668}}, 420 (2008).

\bibitem{gogny-d1m} 
S. Goriely, S. Hilaire, M. Girod  and S. P\'eru, 
Phys. Rev. Lett. {\bf 102}, 242501 (2009).

\bibitem{gogny} 
J. Decharg\'e and D. Gogny, 
Phys. Rev. C {\bf 21}, 1568 (1980).

\bibitem{BCP-1} M. Baldo, P. Schuck and X. Vi\~nas, Phys. Lett. B {\bf 663}, 390 (2008).

\bibitem{BCP-2} L. M. Robledo, M. Baldo, P. Schuck and X. Vi\~nas, Phys. Rev. C {\bf 77}, 051301 (2008).

\bibitem{BCP-3} L. M. Robledo, M. Baldo, P. Schuck and X. Vi\~nas, Phys. Rev. C {\bf 81}, 034315 (2010).

\bibitem{BCP-4} M. Baldo, L. M. Robledo, P. Schuck and X. Vi\~nas, J. Phys. G: Nucl. Part. Phys. {\bf 37}, 064015 (20110).

\bibitem{Robledo-Bertsch-Q3-1} L. M. Robledo and J. F. Bertsch, Phys. Rev. C {\bf 84}, 054302 (2011).

\bibitem{Robledo15} L. M. Robledo, J. Phys. G: Nucl. Part. Phys. {\bf 42} 055109 (2015)

\bibitem{Rayner_Q2Q3_GCM_2012} R. Rodr\'iguez-Guzm\'an, L. M. Robledo 
and P. Sarriguren, Phys. Rev. C {\bf 86}, 034336 (2012). 

\bibitem{rs} P. Ring and P. Schuck, {\it {The Nuclear Many-Body Problem}} 
(Springer, Berlin-Heidelberg-New York) (1980).

\bibitem{Robledo_2D-GCM_with_Butler} L. M. Robledo and P. A. Butler, Phys. Rev {\bf C} 88, 051302 (2013).


\bibitem{Review_RoToRa_2019} L. M. Robledo, Tom\'as R. Rodr\'iguez 
and R. Rodr\'iguez-Guzm\'an, J. Phys. G: Nucl. Part. Phys. {\bf 46}, 013001 (2019).



\bibitem{ours-PT} R. Rodr\'{\i}guez-Guzm\'an, P. Sarriguren, L.M. Robledo
and J.E. Garc\'ia-Ramos, Phys. Rev. C {\bf 81}, 024310 (2010).
  

\bibitem{ours-Y-Nb-quasi} R. Rodr\'iguez-Guzm\'an, P. Sarriguren and L.M.Robledo, 
                          Phys. Rev. C {\bf 83}, 044307 (2011). 

\bibitem{Rayner-fission-1} R. Rodr\'iguez-Guzm\'an 
and L. M. Robledo,  Phys. Rev. C {\bf{89}}, 054310 (2014).


\bibitem{Rayner-fission-5} R. Rodr\'iguez-Guzm\'an 
and L. M. Robledo, Eur. Phys. J. A {\bf{53}}, 245 (2017).

\bibitem{EXP-DATA} T. Kib\'edi and R. Spear, {\it Atomic Data and Nuclear Data Tables} {\bf 80}, 35 (2002).



\bibitem{rob11}  L.M. Robledo and G.F. Bertsch, 
                 Phys. Rev {\bf C} 84, 014312 (2011).

\bibitem{rod02}   R. R. Rodr\'{\i}guez-Guzm\'an, J. L. Egido, and  L.M. Robledo,
                  Nucl. Phys. {\bf A709}, 201 (2002).

\bibitem{egi04}  J.L Egido and L.M. Robledo, 
                 Lecture Notes in Physics {\bf 641}, 269 (2004).

\bibitem{robledo_presciption-1} L. M. Robledo, Int. J. of Mod. Phys. E {\bf 16}, 337 (2007).

\bibitem{robledo_presciption-2} L. M. Robledo, J. Phys. G: Nucl. Part. Phys. {\bf 37}, 064020 (2010).

		 
\bibitem{har82} K. Hara, A. Hayashi and P.Ring, Nucl. Phys. {\bf A385}, 14 (1982).

\bibitem{bon90} P. Bonche, J. Dobaczewski, H. Flocard, P.-H. Heenen 
and J. Meyer, Nucl. Phys. {\bf A510}, 466 (1990).
		 
		 
\bibitem{LMG-model} L. M. Robledo, Phys. Rev. C {\bf{46}}, 238 (1992).

\bibitem{B_Ro_limitations_ROT} L. M. Robledo and J. F. Bertsch, Phys. Rev. C {\bf 86}, 054306 (2012).

\bibitem{limitations_ROT_Rob_EPJA} L. M. Robledo, Eur. Phys. J. A {\bf{52}}, 300 (2016).

\bibitem{gonzalez18}
C. Gonzalez-Boquera, M. Centelles, X. Vi\~nas and L.M. Robledo,
Phys. Lett. {\bf B779} (2018) 195.

\end{thebibliography}
\end{document}